\documentclass[12pt]{article}

\usepackage{epsfig,latexsym,amsfonts,amsmath,amsthm,amssymb, mathrsfs,amsbsy,multirow,slashed,wasysym,textcomp,subfigure,hyperref,wrapfig,datetime,comment,mathtools,cancel,cite}
\usepackage[font={footnotesize,sl},bf]{caption}
\usepackage{comment}

\def\be{\begin{equation}}
\def\ee{\end{equation}}
\def\bea{\begin{eqnarray}}
\def\eea{\end{eqnarray}}
 \newcommand{\badat}{\begin{alignedat}}
 \newcommand{\eadat}{\end{alignedat}}

\long\def\new#1\endnew{{\bf #1}}		
\long\def\del#1\enddel{}

\def\del{\partial}

\usepackage{xcolor}
\definecolor{oldmauve}{rgb}{0.4, 0.19, 0.28}
\definecolor{pansypurple}{rgb}{0.47, 0.09, 0.29}
\definecolor{burgundy}{rgb}{0.5, 0.0, 0.13}
\definecolor{carminepink}{rgb}{0.92, 0.3, 0.26}
\definecolor{blue(pigment)}{rgb}{0.2, 0.2, 0.6}
\definecolor{darkseagreen}{rgb}{0.56, 0.74, 0.56}
\definecolor{darkspringgreen}{rgb}{0.09, 0.45, 0.27}
\definecolor{ceruleanblue}{rgb}{0.16, 0.32, 0.75}

\begin{document}
\numberwithin{equation}{section} 

\begin{titlepage}
  \thispagestyle{empty}
  
  \begin{center} 
  \vspace*{3cm}
{\LARGE\textbf{Exploring the Kleinian horizons}}

\vskip1cm

   \centerline{Gaston Giribet$^{1}$, Juan Laurnagaray$^{2}$, Bryan Malpartida$^{2}$, Pedro Schmied$^{2}$}
\vskip1cm

{$^1$Center for Cosmology and Particle Physics, Department of Physics, New York University},\\ {\it726 Broadway, New York City, NY10003, USA.}\\
{$^2$Departamento de F\'{\i}sica, FCEN, Universidad de Buenos Aires and IFIBA-CONICET}\\ {\it Ciudad Universitaria, pabell\'on 1, 1428, Buenos Aires, Argentina.}\\

\end{center}

\vskip1cm

\begin{abstract}
Self-dual black holes in $(2,2)$ signature spacetime—Klein space—have recently attracted interest in the context of celestial holography. Motivated by this development, we investigate the structure of spacetime near the horizons of these solutions. Focusing on the self-dual Schwarzschild-Taub-NUT solution, we demonstrate that, near the Kleinian horizons, the geometry exhibits a local infinite-dimensional symmetry generated by supertranslations and superrotations. Establishing this result requires refining and extending earlier analyses of asymptotic symmetries near null surfaces. We formulate the appropriate boundary conditions, derive the infinite-dimensional algebra underlying the local symmetries, and compute the associated Noether charges, finding them to be integrable. Finally, we discuss the connection of our findings to recent observations in the literature regarding self-dual black holes in Klein space, including the diffeomorphism relating static and stationary solutions.

\end{abstract}

\end{titlepage}

\section{Introduction}

In \cite{1}, an interesting study of black holes in Klein space was initiated. There, the authors analyzed the analytic continuation of the Lorentzian Kerr-Taub-NUT geometry to its Kleinian counterpart—a stationary black hole with $(2,2)$ signature—characterized by three parameters: the mass $M$, the NUT charge $N$, and the angular momentum per unit mass $a$. It was shown that the global structure of the Kleinian Kerr-Taub-NUT solution is captured by a toric Penrose diagram that closely parallels that of its Lorentzian analogue. A particularly interesting case arises when the solution is self-dual, i.e., when the mass equals the NUT parameter ($M = N$). In this case, it was demonstrated that the Kerr rotation parameter $a$ can be eliminated via a large diffeomorphism—a phenomenon reminiscent of the Euclidean ($0,4$) signature case, with no direct analogue in Lorentzian ($1,3$) signature.

The analytic continuation from $(1,3)$ to $(2,2)$ signature is naturally related to the analytic structure of the scattering matrix in asymptotically flat spacetimes. In \cite{1}, it was shown that the geometry of linearized Kleinian black holes is encoded in the three-point scattering amplitudes of gravitons and massive spinning particles in $(2,2)$ signature. This contrasts sharply with the Lorentzian case, where such amplitudes vanish, highlighting a more direct connection between gravitational backgrounds with $(2,2)$ signature and scattering matrix problems.

Subsequent work further explored the role of Kleinian self-dual black holes in celestial holography. In \cite{4}, the two-dimensional states associated with linearized spinning self-dual black holes in $(2+2)$ dimensions were constructed. These states are described by conformal primaries defined on the celestial torus, and coherent states carrying an infinite tower of $w_{1+\infty}$ charges were identified, establishing links with recent advances in scattering amplitudes and celestial holography in curved backgrounds.

Scattering processes in the self-dual Taub-NUT space have also been studied in \cite{5}, while scattering amplitudes for black holes carrying NUT charge were discussed in \cite{3}. In \cite{2}, the hidden large symmetries near black hole horizons were examined through analytic continuation to $(2,2)$ signature, revealing that the near-horizon region of extremal black holes becomes a Kleinian static solution with mass $M$ and NUT charge $N$. Focusing on the self-dual case ($M = N$) enabled a detailed analysis of approximate and exact near-horizon symmetries. Further aspects of Kleinian black holes have been explored in \cite{6,8,9}, and self-dual solutions have been investigated in \cite{7,10,11}; see also \cite{Kol:2019nkc, Kol:2020ucd} and references thereof.

Motivated by these developments, in this work we investigate the spacetime structure near the horizons of self-dual Schwarzschild-Taub-NUT solutions in Klein space. We prove that, near the Kleinian horizons, the self-dual geometry exhibits a local infinite-dimensional symmetry generated by a semi-direct sum of the supertranslation algebra \cite{Hawking:2015qqa} and the Witt algebra that governs superrotations \cite{12,Hawking:2016msc}. Establishing this result requires refining and extending previous analyses of asymptotic symmetries near null surfaces, in particular by generalizing the fall-off conditions introduced in \cite{13}. Near-horizon symmetries similar to those discussed here have appeared in various contexts \cite{Hawking:2015qqa, 12, Hawking:2016msc, 13, Donnay:2018ckb,14,15,16,Lucho2, Donnay:2020yxw}; see also \cite{Grumiller:2019fmp}. The geometry of spaces with NUT charge has been analyzed in \cite{17}, and related boundary conditions were recently proposed in \cite{Akbar:2025wcq}. However, the case of self-dual black hole horizons in Klein space differs fundamentally from those previously studied.

After briefly reviewing Kleinian black hole geometry in Section 2, we define in Section 3 the boundary conditions appropriate for describing the near-horizon region and derive the infinite-dimensional algebra governing the local symmetries. In Section 4, we compute the associated Noether charges, finding them to be integrable and expressible in terms of the exponential integral function. In Section 5, we discuss the infinite-dimensional horizon symmetries and their connection to recent observations regarding the relationship between static and stationary solutions in Klein space. Section 6 contains our conclusions.

\section{Kleinian black holes}

The usual Taub-NUT metric in Lorentzian ($1,3$) signature reads
\begin{equation}
    ds_{\text{L}}^2 = - f(r) (dt - 2N \cos\theta d\phi)^2 + \frac{dr^2}{f(r)} + (r^2 + N^2)(d\theta^2 + \sin^2 \theta d\phi^2) \, ,\label{una}
\end{equation}
with
\begin{equation}
    \begin{aligned}
        f(r)  = \frac{r^2 - 2 M r - N^2}{r^2 + N^2} = \frac{(r - r_+)(r - r_-)}{r^2 + N^2} \, .
        \end{aligned}\label{dos}
\end{equation}
The metric has two horizons, located at
\begin{equation}
    \begin{aligned}
r_\pm = M \pm \sqrt{M^2 + N^2} \, .
    \end{aligned}
\end{equation}
Here, the Newton constant is set to 1.

It is well known that the above solution presents a Misner string-type singularity at $\theta =0$ and $\theta =\frac{\pi}{2}$. The singularity in either one of the poles can be eliminated by performing the change
\begin{equation}
t\to t\pm 2N\cos \theta \, ;\label{chango}
\end{equation}
the sign $+$ corresponds to the solution without singularity in the northern hemisphere ($0\leq \theta \leq  \frac{\pi}{2}$), while the sign $-$ corresponds to the solution without singularity in the southern hemisphere ($\frac{\pi}{2}\geq \theta \geq  {\pi}$). The change (\ref{chango}), however, introduces a priodicity condition for $t$, with a period $4\pi N$.

The Taub-NUT solution (\ref{una}) can be regarded as the gravitational analog of a magnetic monopole, and the periodicity in $t$ can be thought of as the gravitational analog of the Dirac quantization condition, with the Misner string corresponding to the Dirac string, and $N$ being the gravitomagnetic charge. The space constructed in this way turns out to be spatially homogeneous and of topology $\mathbb{R}\times S^3$. It has spatial isometry $SO(3)$, and thus is spherically symmetric. However, it contains closed time-like curves. The Euclidean, ($0,4$) signature version of it represents the antonomasia of a gravitational instanton.

Here, we will be mostly concerned with the ($2,2$) signature version of the geometry (\ref{una})-(\ref{dos}). Spaces with ($2,2$) signature, often referred to as Klein spaces after Felix Klein's pioneering work in group theory and projective geometry, are frequently considered in the study of scattering amplitudes, since they allow a wider variety of couplings that are nonzero on-shell. In the context of celestial holography, it was shown in \cite{1} that the Schwarzschild-Taub-NUT metric with ($2,2$) signature admits an interpretation in terms of scattering amplitudes in asymptotically $\mathbb{R}^{2,2}$ spacetime. These spaces with Kleinian signature also play an important role in the theory of twistors and conformal geometry in four dimensions; see also \cite{Alty:1993xn, Alty:1993xn, Barrett:1993yn} for pioneering works. Our main motivation to study this comes from recent applications to celestial holography. 

The Kleinian version of the self-dual Taub-NUT geometry is obtained from (\ref{una})-(\ref{dos}) by performing the following complex coordinate redefinition
\begin{equation}
    \begin{aligned}
        t  \to i \, t \, , \ \ \  \theta  \to i \, \theta \, , \ \ \
        N  \to i \, N \, .
    \end{aligned}
\end{equation}
The parameter $N$ must also be Wick-rotated in order to satisfy the reality condition for the metric. The metric of the Kleinian Taub-NUT spacetime obtained in this way takes the form
\begin{equation}
    ds^2_{\text{K}} = f(r) (dt - 2N \cosh\theta d\phi)^2 + \frac{dr^2}{f(r)} - (r^2 - N^2)(d\theta^2 + \sinh^2 \theta d\phi^2) \, ,\label{unados2}
\end{equation}
where now
\begin{equation}
    \begin{aligned}
        f(r)  = \frac{r^2 - 2 M r + N^2}{r^2 - N^2} = \frac{(r - r_+)(r - r_-)}{r^2 - N^2} \, 
    \end{aligned}\label{unados}
\end{equation}
with
\begin{equation}
    \begin{aligned}
 r_\pm = M \pm \sqrt{M^2 - N^2} \, .
    \end{aligned}\label{dosuna}
\end{equation}

Hypersurfaces at $r=r_{\pm}$ are null. They are the natural continuation of the two horizons that the Lorentzian solution has; so we will abuse terminology and refer to them as ``Kleinian horizons''.

Asking the metric (\ref{unados2})-(\ref{unados}) to be smooth on the surfaces $\theta = 0$ and $r = r_+$, in the generic case $M \neq N$, yields the periodicity conditions
\begin{equation}
    \begin{aligned}
        (t, \phi) & \sim (t - 4 \pi N, \phi + 2 \pi), \quad \text{from} \, \, \theta = 0 \\
        (t, \phi) & \sim (t + 4 \pi \frac{r_+^2 - N^2}{r_+ - r_-}, \phi + 2 \pi), \quad \text{from} \, \, r = r_+.
    \end{aligned}
\end{equation}

A very special case is the self-dual configuration, which corresponds to $M = \pm N$. In this case,
\begin{equation}
f(r) = \dfrac{r - M}{r + M}, 
\end{equation}
with a single non-degenerate horizon at $r_{+}=M$. The metric thus becomes
\begin{equation}\label{FFFF}
        ds^2_{\text{K}} = \frac{r - M}{r + M} (dt - 2M \cosh\theta d\phi)^2 + \frac{r + M}{r - M} dr^2 - (r^2 - M^2)(d\theta^2 + \sinh^2\theta d\phi^2) \, .
\end{equation}
The periodicity conditions for this case change. Demanding the ($t,r$) plane to be free of conical singularity around $r=M$ demands $t$ to have a period $8\pi M$, while $\phi $ continues to be periodic in $2\pi$; see below. The potential singularity at $\theta = 0$ demands a period $4\pi M$ for $t$. Summarizing, the following compatibility condition in the plane ($t,\phi $) is obtained 
\begin{equation}
    \begin{aligned}
        (t, \phi)  \sim (t + 4 \pi M, \phi + 2 \pi), 
    \end{aligned}
\end{equation}
with the coordinate ranges $r\in (-M, \infty)$, $\theta \in (0, \infty)$. That is to say, there are two space-like coordinates, $t$ and $r$, with $t$ periodic and $r$ being a non-compact semi-infinite direction. And there are two time-like coordinates, $\phi$ and $\theta$, with $\phi $ periodic and $\theta$ being semi-infinite.
This is exactly the spacetime we will be involved with. Some properties of it are in order:
\begin{itemize}
\item The Kleinian geometry (\ref{FFFF}) is Ricci flat but exhibits a curvature singularity at $r=-M$. This can be seen by computing the Kretschmann scalar $R_{\mu\nu\alpha\beta }R^{\mu\nu\alpha\beta }={96M^2}/{(r+M)^6}$.

\item It is a self-dual solution to Einstein equations; i.e., its Riemann tensor satisfies the relation $\text{sign} (N)\,R_{\mu\nu\alpha\beta }={}^*R_{\mu\nu\alpha\beta }\equiv\frac 12 \varepsilon_{\mu\nu\sigma\eta}R^{\sigma \eta }_{\ \ \alpha \beta }$, with $N=\text{sign} (N) M$. 

\item The geometry can be thought of as a fiber over the 3-dimensional Warped-Anti de Sitter space (WAdS$_3$) \cite{WAdS3}, which is closely related to the Near-Horizon-Extremal-Kerr (NHEK) geometry \cite{Bardeen, sandinista} that plays an important role in the Kerr/CFT correspondence \cite{KerrCFT}. The isometry group of the Kleinian self-dual Taub-NUT is generated by four Killing vectors and is that of WAdS$_3$, i.e., $SL(2,\mathbb{R})\times U(1)$.

\item The space (\ref{FFFF}) is asymptotically locally $\mathbb{R}^{2,2}$. Coordinates $t$ and $\phi$ correspond to two cycles for $M>0$. This yields a toric Penrose diagram whose asymptotic behavior is solidary with the celestial torus at null infinity; see \cite{1}. The condition $\theta >0$ is needed for the cycle not to degenerate, which would actually happen if $\theta = 0$. The $d\phi =0$ sections of the space in the large $r$ limit behave like $ds^2_2\simeq dt^2+dr^2-r^2d\theta^2$, which is a Lorentzian 3-dimensional locally $S^1\times\mathbb{R}^{1,1}$ flat space, with one periodic and two semi-infinite directions. The $dr=d\theta =0$ sections in the large $r$ limit for $\theta >0$ behave like the $S^1\times S^1$ torus $ds^2_2\simeq dt^2+r^2\sinh^2\theta d\phi^2+...$, where the ellipsis stand for $\mathcal{O}(r^0)$ subleading terms in the $g_{\phi \phi}$ and $g_{\phi t}$ components. For $\theta =0$, in contrast, the behavior is quite different, as it collapses to the form $ (dt-2M \cosh \theta d \theta)^2$.

\item The $dt=dr=0$ sections of the space (\ref{FFFF}), when expanding near $\rho \equiv r-M\simeq 0$, yield the 2-dimensional metric $ds^2_2\simeq 2M\rho (d\phi^2-d\theta^2)$, which has $S^1\times \mathbb{R}_+$ constant-$\rho$ foliations. On the other hand, the $d\theta =d\phi=0$ sections, near the region $r \simeq M$, yield the characteristic Euclidean semi-infinite cigar form $ds^2_2\simeq H{(\rho )}dt^2+d\rho^2/H{(\rho)}$ with $H(\rho)={\rho}/({2M})$, which demands the periodicity $\beta = 4\pi /\partial_{\rho}H(\rho )= 8\pi M$ in $t$ in order to avoid conical singularity at the tip $\rho =0$. The topology of the ($t,\rho $) plane is that of $\mathbb{R}^2$ with asymptotic behavior $S^1\times \mathbb{R}$. Last, the $dt=d\phi=0$ sections yield the locally flat 2-dimensional metric $ds^2_2\simeq 8M( dR^2-R^2 d({\theta}/{2})^2)$, with $R^2=\rho=r-M$.

\item It was explicitly shown in \cite{1} that the self-dual static metric (\ref{FFFF}) is diffeomorphic to a stationary, seemingly spinning version of it; see (\ref{quera1}) below. A particular case of this is the well-known relation between Minkowski space and the massless limit of the over-spinning Kerr solution.

\item The metric (\ref{unados2})-(\ref{unados}) also exhibits symmetry under the CPT type $\mathbb{Z}_2$ transformations $(N;t, \phi )\to (-N;t, -\phi)$, $(N;t,\phi )\to (-N;-t,\phi )$, and $(N; t ,\phi )\to (N; -t,-\phi)$. The stationary metric has an additional parity-odd Kerr-type parameter $a$.

\item  The null surface at $r=M$ is geodesically complete and so regular, and it corresponds to the analytic continuation of the non-degenerated event horizon that the Lorentzian geometry (\ref{una}) exhibits at the self-dual points $N=\pm M$. Below, we will prove that, near these null surfaces, the Kleinian geometry (\ref{FFFF}) exhibits infinite-dimensional symmetries. 
\end{itemize}

In preparation for the analysis of near-horizon symmetries, we can write the self-dual metric (\ref{FFFF}) in a suitable coordinate system. Introducing a radial coordinate $\rho = r-M $ and an advance time $ v = 2 M \theta + 2 M \log{{\rho}}$, the $g_{\rho\rho}$ component of the metric is found to vanish, and thus the metric takes the form
\begin{align}\label{G213}
ds^2_{\text{K}} &=  - \frac{\rho}{2M} (\rho + 2 M) \left(\frac{dv^2}{2M} - \frac{2}{\rho} dvd\rho\right) -  \frac{4M\rho}{\rho + 2 M} \left(\frac{M}{\rho} e^{\frac{v}{2M}} + \frac{\rho}{4M} e^{-\frac{v}{2M}} \right)\,dt d\phi\nonumber\\
& + \frac{\rho}{\rho + 2 M} dt^2 - \left(\frac{\rho}{\rho + 2 M} \left(\rho^2 + 4M \rho\right) \left(\frac{M}{\rho} e^{\frac{v}{2M}} - \frac{\rho}{4M} e^{-\frac{v}{2M}} \right)^2 - \rho (\rho + 2 M)\right)\,d\phi^2\,.
\end{align}
Then, the non-zero components are
\begin{align}
    g_{vv} & = - \frac{1}{2M} \rho + \mathcal{O}\left(\rho^2\right)\nonumber \\
    g_{v\rho} & = g_{\rho v} =  1 + \frac{1}{2M} \rho + \mathcal{O}\left(\rho^2\right)\nonumber \\
    g_{tt} & = \frac{1}{2M} \rho + \mathcal{O}\left(\rho^2\right) \\
    g_{t\phi} & = g_{\phi t}= -M e^{\frac{v}{2M}} + \mathcal{O}\left(\rho\right)\nonumber \\
    g_{\phi\phi} & = -2M^2 e^{\frac{v}{M}} + \mathcal{O}\left(\rho\right)\nonumber \,.
\end{align}

Note that the induced metric on the self-dual Kleinian Taub-NUT horizon will now depend on the advance time coordinate $v$. In fact, the induced (1+1)-dimensional metric on the constant-$v$ foliations of the horizon, $H$, is
\begin{align}
    ds^2_H = -2M  e^{\frac{v}{2M}} dt d\phi - 2 M^2 e^{\frac{v}{M}} d\phi^2 \equiv \Omega_{AB}\, dx^Adx^B\, ,
\end{align}
where we denoted $x^1=t$, $x^2=\phi$, $A,B=1,2$; which means $\Omega_{11}=0$, $\Omega_{12}=\Omega_{21}=-Me^{\frac{v}{2M}}$, $\Omega_{22}=-2M^2e^{\frac{v}{M}}$. This explicit dependence of $v$ will be important to understand the functional form of the charges below.

\section{Near-horizon symmetry}

We are interested in studying the asymptotic symmetries of the Kleinian geometry near the horizon. To do so, we need to describe the metric close to null surfaces. We will choose a suitable set of coordinates such that the distance from the horizon is measured by a radial coordinate $\rho$, with $v$ being the advanced time on $H$, and $x^A$ describing the transversal $(1+1)$-dimensional spacetime. Typically, the transversal constant-$v$ sections of the horizon are Euclidean; in the Kleinian case these sections are replaced by Lorentzian surfaces.

Following the procedure described in \cite{13}, it is found that the metric on the null surface must obey
\begin{equation}\label{eq:cond-cont}
g_{\rho\rho} = 0,\ \ \ g_{\rho A}=0,\ \  \ g_{\rho v} = 1.
\end{equation}
In order to satisfy \eqref{eq:cond-cont} we assume that the corresponding component of the metric admits the expansion
\begin{equation}\label{eq:ccs}
g_{\rho\rho} = \alpha \rho + \mathcal{O}(\rho^2), \quad g_{A\rho} =  \beta_A \rho +\mathcal{O}(\rho^2), \quad g_{v\rho} = 1 + \gamma \rho + \mathcal{O}(\rho^2),
\end{equation}
with $\alpha, \beta_A, \gamma$ independent of $\rho$. At the Kleinian horizon, which is located at $\rho =0$, the prescription \eqref{eq:ccs} clearly satisfies the conditions \eqref{eq:cond-cont}. Notice that (\ref{eq:ccs}) is more relaxed than the near-horizon asymptotic conditions considered in \cite{12} and \cite{13}. The latter correspond to the particular case $\alpha =\beta_A=\gamma = 0$. Weak asymptotic conditions similar (\ref{eq:ccs}) were recently considered in \cite{Akbar:2025wcq}.

For the other components of the metric we consider the expansion
\begin{equation}
g_{vv} = -2\kappa \rho + \mathcal{O}(\rho^2),\quad g_{vA} = \theta_A\rho  +\mathcal{O}(\rho^2),\quad g_{AB} = \Omega_{AB} + \omega_{AB}\rho+\mathcal{O}(\rho^2). \label{eq:ccsxxx} 
\end{equation}
Functions $\alpha$, $\beta_A$, $\gamma$, $\kappa$, $\theta_A$, $\Omega_{AB}$ and $\omega_{AB}$ are assumed to depend on $v$ and $x^A$, but are independent of $\rho$. $\Omega_{AB}$ is assumed to be non-degenerate.

Given the above boundary conditions we solve the asymptotic Killing equation for the asymptotic vector $\xi$. That is to say, we look for the $\xi$ that preserve the boundary conditions (\ref{eq:ccs})-(\ref{eq:ccsxxx}). This amounts to solve
\begin{equation}\label{eq:c.cs}
\mathcal{L}_\xi g_{\rho\rho} = \mathcal{O}(\rho^2), \quad \mathcal{L}_\xi g_{\rho v} = \mathcal{O}(\rho^2), \quad \mathcal{L}_\xi g_{\rho A} = \mathcal{O}(\rho^2),
\end{equation}
where $\mathcal{O}(\rho^n)$ stands for a function of the form $F(v,x^A) \rho^{n}$ or subleading near $\rho \simeq 0$. From this conditions, we obtain the expansion in powers of $\rho$ for each component of $\xi$; namely
\begin{equation}\label{eq:killing}
    \begin{split}
            \xi^v & = f-\frac{Z\alpha}{2}\rho+\mathcal{O}(\rho^2)\, ,\\
            \xi^A & = Y^A - \Omega^{AB}\left(\partial_B f+Z \beta_B\right)\rho+\mathcal{O}(\rho^2)\, ,\\
            \xi^\rho & = Z - (Z\gamma +\partial_v f)\rho + \mathcal{O}(\rho^2) \, .
    \end{split}
\end{equation}
where $f$, $Z$ and $Y^A$ are functions of $v$ and $x^A$ that do not depend on $\rho$. Preserving the boundary conditions \eqref{eq:ccs} demands $\mathcal{L}_\xi g_{ij} = \delta g_{ij}$ with $i$ and $j$ running over $v$ and the {\it angular} coordinates $x^A$. This is expressed in terms of the components of the metric as follows
\begin{align}
\mathcal{L}_\xi g_{vv} & = -2  \delta\kappa \, \rho+ \delta g_{vv}^{(2)}\, \rho^2+\mathcal{O}(\rho^3),\\
\mathcal{L}_\xi g_{vA} & =  \delta\theta_A\, \rho + \delta g_{vA}^{(2)}\, \rho^2+\mathcal{O}(\rho^3),\\
\mathcal{L}_\xi g_{AB} & = \delta \Omega_{AB} +  \delta\omega_{AB} \, \rho + \delta g_{AB}^{(2)}\, \rho^2+\mathcal{O}(\rho^3).
\end{align}
From this, we get that $Z$ and $Y^A$ are obtained by solving the equations
\begin{equation}\label{eq:YyZ}
\partial_v Z -\kappa Z = 0,\quad Z \theta_A + \partial_A Z + \Omega_{AB}\partial_vY^B = 0\, .
\end{equation}
This yields the following variations of the metric functions
\begin{equation}\label{eq:variaciones}
\begin{split}
\delta \kappa & = Y^A\partial_A\kappa + \partial_v(\kappa f)+\partial_v^2f - \theta_A\partial_v Y^A + Z \left(\partial_v \gamma - g_{vv}^{(2)}-\kappa \gamma \right),\\
\delta \theta_A & = \theta_B\partial_A Y^B+Y^B\partial_B\theta_A+f\partial_v\theta_A-2\kappa\partial_Af-2\partial_A\partial_v f+ \Omega^{BC}\partial_v(\Omega_{AB})\partial_C f+\omega_{AB}\partial_v Y^B\\
& \quad\,+Z\left(2 g_{vA}^{(2)}-\partial_A \gamma - \theta_A \gamma-\Omega_{AB}\partial_v(\Omega^{BC}\beta_C)\right),\\
\delta \Omega_{AB} & = Z \omega_{AB}+f \partial_v\Omega_{AB}+\mathcal{L}_Y \Omega_{AB},\\
\delta \omega_{AB} & = -\omega_{AB}\partial_v f + f\partial_v \omega_{AB}+ \mathcal{L}_Y \omega_{AB}+\theta_A \partial_B f+ \theta_B\partial_A f-2\nabla_A\nabla_B f+2 Z g_{AB}^{(2)}-Z\gamma\omega_{AB}\\
& \quad\, -\frac{Z\alpha}{2}\partial_v\Omega_{AB}+\beta_A\partial_B Z+\beta_B\partial_A Z-\Omega_{AC}\partial_B(Z\Omega^{CD}\beta_D)-\Omega_{BC}\partial_A(Z\Omega^{CD}\beta_D).
\end{split}
\end{equation}
This generalizes the results of \cite{13} and reduces to it when $\alpha=\beta_A=\gamma = 0$.

We will assume that the prescription of the asymptotic boundary condition near $H$ is state-independent. This implies that the asymptotic killing vectors do not depend on the metric functions themselves. Then, from \eqref{eq:YyZ} we get
\begin{equation}\label{eq:ccZeY}
Z = 0, \quad \partial_vY^B = 0.
\end{equation}
From this assumption, equations \eqref{eq:killing} and \eqref{eq:variaciones} became
\begin{equation}\label{eq:killing-simples}
    \begin{split}
            \xi^v & = f+\mathcal{O}(\rho^2),\\
            \xi^A & = Y^A - \Omega^{AB}\partial_B f\rho+\mathcal{O}(\rho^2),\\
            \xi^\rho & =  - \partial_v f\rho + \mathcal{O}(\rho^2).
    \end{split}
\end{equation}
and then we obtain
\begin{equation}\label{eq:variaciones-simples}
    \begin{split}
        \delta \kappa & = Y^A\partial_A\kappa + \partial_v(\kappa f)+\partial_v^2f,\\
        \delta \theta_A & = \mathcal{L}_{Y}\theta_A+f\partial_v\theta_A-2\kappa\partial_Af-2\partial_A\partial_v f+\Omega^{BC}\partial_v(\Omega_{AB})\partial_C f\\
        \delta \Omega_{AB} & = f \partial_v\Omega_{AB}+\mathcal{L}_Y \Omega_{AB},\\
        \delta \omega_{AB} & = -\omega_{AB}\partial_v f + f\partial_v \omega_{AB}+ \mathcal{L}_Y \omega_{AB}+\theta_A \partial_B f+ \theta_B\partial_A f-2\nabla_A\nabla_B f.
    \end{split}
\end{equation}

By demanding the asymptotic behavior \eqref{eq:ccZeY} on \eqref{eq:killing}, we recover the Killing vectors derived in \cite{13}. The modified Lie brackets of these asymptotic diffeomorphisms is defined as 
\begin{equation}\label{algebrotaA}
\{\xi_1,\xi_2\}\, \equiv \, \mathcal{L}_{\xi_1}\xi_2-\delta_{\xi_1}\xi_2+\delta_{\xi_2}\xi_1,
\end{equation}
which closes the algebra of vectors
\begin{equation}
\{\xi_1(f,Y^A_1),\xi_2(f,Y^A_2)\} = \xi_{12}(f_{12},Y_{12}^A)
\end{equation}
where
\begin{equation}\label{algebrotaB}
f_{12} = f_1\partial_v f_2 -f_2\partial_vf_1+Y_1^A\partial_A f_2-Y_2^A\partial_A f_1,\; Y_{12}^A = Y_1^B\partial_B Y_2^A-Y_2^B\partial_B Y_1^A.
\end{equation}

Remarkably, we obtained a consistent closed algebra having allowed for a more general behavior on the metric components $g_{\rho \mu}$, which now admit non-trivial $\mathcal{O}(\rho)$ terms. This algebra contains supertranslations ($f$) in semi-direct sum with two copies of the Witt algebra ($Y_A$).

Now, we can apply this technology to the solution of Einstein equations we are interested in, namely the self-dual Kleinian Taub-NUT solution (\ref{FFFF}). Performing, as in (\ref{G213}), the change of coordinates
\begin{equation}
    r = \rho + M, \qquad \theta = \frac{v}{2M} - \log{\rho} \, ,
\end{equation}
which suffices to realize the boundary conditions introduces above, we are ready to express the metric in coordinates $(t, \rho, v, \phi)$, with $v$ being the advanced time on the furface $r=M$, the latter being mapped into $\rho =0$. We also have $x^1=t, x^2=\phi$. The asymptotic behavior thus reads
\begin{equation}\label{eq:lametrica}
    ds_{\text{K}}^2 =  \frac{1}{2M}\rho \, dv^2 +2(1+\frac{1}{2M}\rho) \, dvd\rho +\frac{1}{2M}\rho \,  dt^2-2 M e^{v/2M}dtd\phi-2M^2 e^{v/M}d\phi^2 + \mathcal O(\rho^2) \, .
\end{equation}
Note that from the above expression we can read the Kleinian analog to the surface gravity from the $g_{vv}^{(1)}$ term, namely $\kappa = \frac{1}{4M}$. This agrees with the analysis of the periodicity of the coordinate $\theta$ around $\rho =0$. In fact, defining the imaginary time $\tau \equiv 2M\, i\theta$, we find that the absence of conical singularity around $r=M$ in the ($\theta , r$) plane requires $i\theta $ to have a period $4\pi $, i.e., $\tau $ to have a period $\beta = \frac{2\pi}{\kappa} = 8\pi M$.

The form (\ref{eq:lametrica}) is the very proof that, in the near-horizon region, the Kleinian self-dual Schwarzschild-Taub-NUT solution exhibits the infinite-dimensional symmetry generated by the algebra (\ref{algebrotaA})-(\ref{algebrotaB}). In the next section, we will derive the Noether charges associated to these infinite symmetry transformations.

\section{Noether charges}

Without additional assumptions about the behavior of the fields, we are able to compute the charge variation. Starting from expression (24) in \cite{13}, we get
\begin{equation}
    \begin{aligned}
        \cancel \delta Q [\xi] & =\frac{1}{16\pi }\int \left(d^2x\right)_{\mu\nu}\sqrt{-g}\bigg[ - \xi^\nu \nabla^\mu h + \xi^\nu\nabla_\sigma h^{\mu\sigma} + \xi_\sigma\nabla^\nu h^{\mu\sigma} \\
        & \hspace{110pt} + \frac{1}{2}h\nabla^\nu \xi^\mu +\frac{1}{2}h^{\nu\sigma}(\nabla^\mu\xi_\sigma - \nabla_\sigma\xi^\mu) - (\mu  \leftrightarrow \nu) \bigg] \, .
    \end{aligned}
    \label{eq:barnich_charge}
\end{equation}
where $h_{\mu\nu}$ is the metric variation ${\delta g_{\mu\nu}}$, and $\xi$ is an asymptotic Killing vector. $(d^2x)_{\mu\nu}$ stands for the component $\mu \nu $ of the integration measure written as a 2-form. The symbol $\cancel \delta$ means that the charge is in general non-integrable.

Unlike previous works in which the choice $\delta \kappa = 0$ was made, here we will not make any assumption about such functional variation. Then, for $Y^A = 0$, from the first equation of \eqref{eq:killing-simples} we have the relation
\begin{equation}
    \delta \kappa = \partial_v \left( \kappa f + \partial_v f \right) \, .
    \label{eq:kappa_variation}
\end{equation}

In the specific case we are dealing with, by defining $h_{\mu \nu} = \delta M{\partial_M g_{\mu \nu}} $ we can compute each term in \eqref{eq:variaciones-simples} explicitly; namely
\begin{equation}
    \begin{split}
        & \xi^\nu \nabla^\mu h = \frac{1}{M} \, e^{v/2M} f \, \delta M  \, ,\\
        & \xi^\nu \nabla_\sigma h^{\mu\sigma} = \frac{v}{4 M^2} \, e^{v/2M} f \, \delta M \, , \\
        & \xi_\sigma \nabla^\nu h^{\mu\sigma} = \frac{1}{2 M} e^{v/2M} f \, \delta M \, ,\\
        &  h \nabla^\nu \xi^\mu = \frac{v - 2M}{2 M^2} e^{v/2M} \left( f + 4 M \partial_v f \right) \, \delta M \, ,\\
        & h^{\nu\sigma}\nabla^\mu \xi_\sigma = 0 \\
        & h^{\nu\sigma}\nabla_\sigma \xi^\mu = 0
    \end{split}
\end{equation}
where $f$ is a function of $t, \phi$ and $v$. Taking all this into account, we obtain the supertranslation charge variation
\begin{equation}
    \delta Q[\xi] = \frac{1}{16\pi }\int_{H_v} d^2x  \,  \left(-f(x,v) + (v - 2M) \, \partial_v f(x,v)  \right) \,  \frac{e^{\frac{v}{2M}}}{M} \, \delta M \, .
    \label{eq:charge_variation_final}
\end{equation}
with $x^A=\{t,\phi\}$, $A=1,2$, $d^2x=dtd\phi$; ${H_v}$ refers to the constant-$v$ sections of $H$. This charge is integrable and, provided $f$ is chosen adequately, it is finite. The expression for the charge $Q$ can easily be written in terms of the exponential integral function
\begin{equation}
\text{Ei}(z)=\int^{\, z}_{-\infty}\, \frac{d\tau}{\tau}\,e^{\tau}\, ;
\end{equation}
although its form is not particularly illuminating. In computing $Q$ one typically assumes that the function $f$ is independent of $M$. However, one can also consider a case such as $f(v)=e^{-\frac{v}{2M}}$, which yields a charge that is linear in $v$ and reminiscent of the Wald entropy charge, with $\partial_v Q=\frac{\text{Area}}{4}\frac{\kappa}{2\pi } $.

Note that the expression (\ref{eq:charge_variation_final}) for $\delta Q$ happens to be $v$-dependent, and so $Q$ is not conserved. This is an aspect inherited from the fact that we are integrating over a surface that explicitly depends on $v$. In other words, the metric induced on the constant-$v$ slices of $H$ is, in turn, $v$-dependent. The appearance of non-conserved charges in the context of asymptotic symmetries in asymptotically flat spacetimes is not new. In the case of BMS symmetries and their generalizations, this non-conservation is associated with outflows, which can be interpreted as gravitational radiation. In the case of the spaces considered here, the interpretation is considerably less clear. The inherent difficulty of conceiving solutions in Klein space, especially when it involves null surfaces that are the analytic continuation of event horizons, makes the task of coming up with an interpretation of the charges and their dynamics very complicated. We hope to return to this problem later. In this work, we limit ourselves to analyzing the geometric aspects of the region close to the Kleinian horizons and, with this, demonstrate the existence of infinite-dimensional symmetries there.



An alternative computation of the charge can be performed by using the expression (25) in \cite{13}, where \eqref{eq:barnich_charge} is already evaluated on the specific configuration of the self-dual Taub-NUT geometry \eqref{eq:cond-cont}. This yields
\begin{equation}
    \begin{aligned}
        \cancel \delta Q [f, Y^A] & = \frac{1}{16 \pi }\int_{H_v} d^2x \, [ 2 \, f \, \kappa \, \delta  \sqrt{\hat \Omega} - 2 \partial_v f \delta \sqrt{\hat \Omega} + 2 f \partial_v \delta \sqrt{\hat \Omega} \\
        & \qquad \qquad - \frac{1}{2} f \sqrt{\hat \Omega} (\Omega^{AB} \Omega^{CD} - \Omega^{AC} \Omega^{BD}) \partial_v \Omega_{CD} \delta \Omega_{AB} + Y^A \delta (\theta_A \sqrt{\Omega})] \, ,
    \end{aligned}
    \label{eq:general_charge_variation}
\end{equation}
where $\hat{\Omega}=-\text{det}\,\Omega_{AB}=M^2e^{\frac{v}{M}}$. For our metric, and taking $Y^A = 0$, this yields
\begin{equation}
    \delta\sqrt{\hat \Omega} = \frac{e^{\frac{v}{2M}}}{2M} (2M -v ) \, \delta M \, , \quad  \, \delta(\theta_A \sqrt{\hat{\Omega}}) = 0 \, .
\end{equation}
Using that
\begin{equation*}
    \sqrt{\hat \Omega} \, (\Omega^{AB} \Omega^{CD} - \Omega^{AC} \Omega^{BD}) \partial_v \Omega_{CD} \delta \Omega_{AB} = \frac{1}{2M^2} e^{\frac{v}{2M}} (2M - v) \, ,
\end{equation*}
and
\begin{equation*}
    \partial_v \delta \sqrt{\hat \Omega} = - \frac{v}{4 M^2} e^{\frac{v}{2M}} \delta M \, ,
\end{equation*}
we get
\begin{equation}
    \begin{aligned}
        \delta Q [f]  =  -\frac{1}{16 \pi }\int_{H_v} d^2x \bigg( f (x,v) + (2M -v )\partial_v f(x,v) \bigg) \frac{e^{\frac{v}{2M}}}{M} \delta M \, ,
    \end{aligned}
\end{equation}
which exactly matches (\ref{eq:charge_variation_final}).

\section{On the spinning geometry}

In \cite{1}, the authors presented a set of coordinates that maps the self-dual Kleinian Schwarzschild-Taub-NUT spacetime into the Kleinian Kerr-Taub-NUT spacetime with $M = \pm N$ for arbitrary Kerr parameter $a$. That is to say, the static and the stationary self-dual Kleinian configurations turn out to be diffeomorphic. Specifically, the authors of \cite{1} showed that the Kleinian Taub-NUT space
\begin{equation}
ds^2_{\text{K}} = \frac{r+M}{r-M}dr^2-(r^2-M^2)(d\theta^2+\sinh^2\theta d\phi^2)+\frac{r-M}{r+M}(dt-2M\cosh\theta d\phi)^2,\label{quera1}
\end{equation}
when the transformation
\begin{align}\label{eq:transformaciondeStrominger}
r = r' +a\cosh\theta'\, , \ \ \ \
\tanh\frac{\theta}{2} = \tanh\frac{\theta'}{2}\, \sqrt{\frac{r'-M-a}{r'-M+a}},
\end{align}
is performed, becomes the self-dual Kleinian Kerr-Taub-NUT space
\begin{equation}
ds^2_{\text{K}} = \Sigma\left(\frac{dr'^2}{\Delta}-d\theta'^2\right)-\frac{\sinh^2\theta'}{\Sigma}(adt- P d\phi)^2+\frac{\Delta}{\Sigma}(dt+Ad\phi)^2
\end{equation}
with
\begin{align}\label{queran}
\Sigma & = r'^2-(M+a\cosh\theta')^2\\
\Delta & = (r'-M)^2-a^2\\
A &= -a\sinh^2\theta'-2M\cosh\theta'\\
P & = r'^2-M^2-a^2.
\end{align}

Then, it is natural to ask whether such transformation corresponds to one of the diffeomorphisms that preserve the asymptotic behavior of the Kleinian self-dual Taub-NUT metric near the horizon, namely those in \eqref{eq:killing-simples}. To see whether this is the case, we can define the new coordinates $r'$ and $\theta'$ as an expansion on the parameter $a$, like $x'^\mu  = x'^\mu_{|a=0} +  a\,\tilde{\xi}^\mu +\cdots.$ The inverse of \eqref{eq:transformaciondeStrominger} for $r>M$ reads
\begin{align}
r'   = \frac{1}{2}\left(M+r+\Theta (r,\theta) 
\right)\, , \ \ \ \
\tanh\frac{\theta'}{2}  = \tanh\frac{\theta}{2}\, \sqrt{\frac{2a-(r-M)\cosh\theta+\Theta (r,\theta) }{(r-M)(1-\cosh\theta )}},
\end{align}
with 
\begin{equation}
\Theta^2(r,\theta ) = 4a^2+(r-N)^2-4a(r-M)\cosh \theta 
\end{equation}
which, expressed in terms of coordinates $\rho$ and $v$, turns out to be generated by the Killing vector
\begin{equation}
\tilde{\xi}^v = - e^{-v/2M},\quad \tilde{\xi}^A = 0,\quad \tilde{\xi}^\rho = -\frac{Me^{v/2M}}{\rho}-\frac{ e^{-v/2M}}{4M}\, \rho\, .
\end{equation}

This vector is singular at $\rho=0$, which is the location of the horizon of the solution with $a=0$. In fact, the diffeomorphism that brings the static self-dual metric into its stationary form changes the location of the horizon, while the latter is preserved by diffeomorphisms (\ref{eq:transformaciondeStrominger}). The presence of an order $\mathcal{O}(\rho^0)$ or greater in the component $\tilde{\xi}^\rho$ makes it clear that Killing vector $\tilde{\xi}$ belongs to a class different from that of the near-horizon asymptotic Killing vectors discussed above. 

\section{Conclusions}

In this paper, we have been involved in the study of geometries with Klein signature; more precisely, with the analytical extension of the self-dual Schwarzschild-Taub-NUT solution to the ($2,2$) signature. These geometries have recently been considered in the framework of celestial holography, where they have been related to three-point scattering amplitudes in asymptotically, locally $\mathbb{R}^{2,2}$ spaces. Here, motivated by these applications, we set out to analyze the geometry near the event horizons of self-dual Kleinian black holes. To do so, we generalized the analysis in \cite{12} of the infinite-dimensional symmetries of the region near the event horizons and extended it to the case of Kleinian horizons of self-dual solutions. The structure of these horizons required to prescribe a set of asymptotic conditions more relaxed than those usually considered, cf. \cite{13}. Despite this, we found that the Kleinian horizons of self-dual black holes exhibit an infinite-dimensional near-horizon symmetry, which is generated by both an algebra of supertranslations and superrotations. Relaxed asymptotic conditions similar to those considered here were independently studied in \cite{Akbar:2025wcq}, and the symmetries of the self-dual black hole were studied for different boundary conditions in \cite{17}. For the case of self-dual Kleinian horizons, we calculated the supertranslation charges associated, which, remarkably, turned out to be integrable. Since the charge calculation is carried out on horizon sections that are dependent on the null time defined over them, the charges inherit an advanced time dependence. We also considered stationary Kerr-Taub-NUT solutions, which in the case of the self-dual solution with signature (2,2) are diffeomorphically equivalent to the static version. We thus compared the diffeomorphism linking both solutions and those asymptotic diffeomorphisms that preserve the near-horizon geometry. The main result of this paper is that, even with the generalization of the near-horizon asymptotic conditions necessary to describe self-dual solution horizons in Klein space, the spacetime geometry near these horizons exhibits an infinite-dimensional symmetry resembling the BMS algebra. The interpretation of the charges associated with these symmetries, however, is conceptually elusive, and it would be necessary to study a dynamic process --for example, similar to the one studied in \cite{Donnay:2018ckb} but adapted to the Kleinian case-- in order to advance in the understanding of the physical meaning of these symmetries and their associated quantities. We plan to address this problem in the near future. 


\section*{Acknowledgements}

The authors thank Luciano Montecchio for discussions in the early stage of this collaboration.

\end{document}